\documentclass[referee]{raa}            

\usepackage{graphicx,times}             

\begin{document}

   \title{NSVS 01286630: A Detached Binary with a Close-in Companion
}

   \volnopage{Vol.0 (200x) No.0, 000--000}      
   \setcounter{page}{1}          

   \author{Bin Zhang
      \inst{1,2,3}
   \and Sheng-bang Qian
      \inst{1,2,3}
   \and Wen-ping Liao
      \inst{1,2,3}
   \and Li-ying Zhu
      \inst{1,2,3}
   \and Ai-jun Dong
      \inst{4}
   \and Qi-jun Zhi
      \inst{4}
   }

   \institute{Yunnan Observatories, Chinese Academy of Sciences (CAS), P. O. Box 110, 650216 Kunming, China; {\it zhangbin@ynao.ac.cn}\\
        \and
             Key Laboratory of the Structure and Evolution of Celestial Objects, Chinese Academy of Sciences, P. O. Box 110, 650216 Kunming, China\\
        \and
             University of Chinese Academy of Sciences, Yuquan Road 19\#, Sijingshang Block, 100049 Beijing, China\\
        \and
             School of Physics and Electronic Science, Guizhou Normal University, 550001 Guiyang, China\\}

   \date{Received~~2009 month day; accepted~~2009~~month day}

\abstract{New photometric observations of NSVS 01286630 were performed and two sets of four-color ($B$, $V$, $R_{c}$, $I_{c}$) light curves (LCs) were obtained.
Using the 2013 version of the Wilson-Devinney $(W-D)$ code, we analyzed these data.
The photometric solutions reveal that NSVS 01286630 is an active detached eclipsing binary (EB) with a high orbit inclination (nearly $90^{\circ}$).
Remarkably, the temperature of the primary component (the hotter star) is higher than the secondary one,
but, the value of mass ratio $q(\frac{M_{2}}{M_{1}})$ for NSVS 0128663 is more than 1, which can be explained that
the surface of the secondary component of NSVS 01286630 is covered with big cool-star-spots.
Based on our new CCD mid-eclipse times and the data published until now, the variations of the mid-eclipse times were reanalyzed in detail using a weighted least-squares method.
It is discovered that the $(O - C)$ diagram of the system shows a cyclic oscillation with a period of 3.61 yr and an amplitude of 0.001 days.
The cyclic variation may be caused by the light travel time effect (LTTE) due to the presence of a third companion,
whose mass we calculated is $M_{3}$sin($i_{3}$)=0.11 $M_\odot$.
The third body may affect the orbital evolution of the central binary system by transferring the angular momentum.
}

   \authorrunning{Bin Zhang et al. }            
   \titlerunning{Detached binary NSVS 01286630}  

   \maketitle

%
%

\section{Introduction}
\label{sect:intro}
A lot of close binaries are formed with a close-in companion, and the
tertiary components play an important role in the formation and evolution of these systems (Wolf et al. 2016; Fabrycky \& Tremaine 2007).
The statistical study using large eclipsing binary (EB) sample looking for evidence of hierarchical triple-star systems is an useful way.
The LAMOST survey released many EB data recently, including some important orbital parameters such as orbital period, effective
temperature, gravitational acceleration and metallicity (Qian et al. 2017).
Using these data, Qian et al. (2017) researched the correlations between orbital period and other parameters and found many EB candidates with a third body.
Another search through the photometric database of Kepler eclipsing binaries (EBs) suggest that at least
20\% of all close binaries have tertiary companions (Gies et al. 2012; Rappaport et al. 2013; Conroy et al. 2014; Borkovits et al. 2015, 2016).
For an EB system, the presence of the tertiary component can cause a cyclical variation of the minimum light times, which can be investigated using the known $(O - C)$ method.
Through analyzing the difference of mid-eclipse times between the observed and computed with a given ephemeris,
we can obtain some orbital parameters of the third body (Liao \& Qian 2010a).
In the other words, we can discover possible multi-systems by searching for the observed $(O - C)$ diagram with periodical features (Wolf et al. 2016).
Using this method, recent years have been reported many successful
examples for the detection of the third body around the close binaries such as
V1104 Her (Liu et al. 2015), V401 Cyg (Zhu et al. 2013), V894 Cyg (Li \& Qian 2014),
KIC 5513861 (Zasche et al. 2015) and KIC 9532219 (Lee et al. 2016).
Specifically, the stable $M$-type contact binary system SDSS J001641-000925 with a close-in star companion was also discovered (Qian et al. 2015a).

NSVS 01286630 (=NSVS 1135262) is first detected as an EB candidate in 2004 (Woniak et al. 2004),
after 3 years, the first paper about this target is published (Coughlin \& Shaw 2007).
The research results from Coughlin \& Shaw (2007) suggest that NSVS 01286630 is a detached EB system with big cool-star-spots
in its polar areas, which implies a strong magnetic activity of this system.
The physical masses and radiuses are also calculated based on assigned masses extrapolated from their temperature.
The preliminary model seems to support the current findings that low-mass stars have a
greater radius and lower effective temperature than the best models predicted (Lacy 1977).
Recently, after analysing the period variations of NSVS 01286630 using the $(O - C)$ method,  
the existence of a cyclic oscillation with an amplitude of 0.00105 days and a period about 1317 days is discovered by Wolf et al. (2016). They thought that this period changes maybe caused due to the presence of a third body,
and according to their fitting parameters, the lowest mass of the third body is calculated as 0.10M$_{\odot}$.
And it is based on that, we monitored this target for several years.

In present paper, first four-color ($B$, $V$, $R_{c}$, $I_{c}$) light curve (LCs) of NSVS 01286630 are obtained and analyzed.
We also used new mid-eclipse times coupled with old data to analyse the $(O-C)$ diagram, which shows a
short-time cycle oscillation because of the presence of the third body.
Based on the photometric analysis and the $(O - C)$ fitting results, the tertiary star, evolution
state and magnetic activity of this system are discussed.
\section{Multi-color CCD Photometric Observations}
\label{sect:Obs}
Two-sets of LCs in $BVR_{c}I_{c}$ bands of NSVS 01286630 were obtained in 2010 and 2011, respectively.
The first set of LCs were observed in 2010 November 2, 3, and the second set of LCs were made on 2011 May 2, 5, 6, June 11, 12, and 20.
A total of more than 1900 CCD images are carried out using the 85 cm telescope at the
Xinglong Station of National Astronomical Observatories of Chinese Academy of Sciences.
This telescope equipped with $1024 \times1024$ PI1024 BFT camera with a standard Johnson-Cousins-Bessel multicolor CCD photometric system,
the effective field of view is 16.5 arcmin by 16.5 arcmin corresponding to a plate scale of 0.97 arcsec pixel$^{-1}$  (Zhou et al. 2009). From 2014 to 2015, the telescope had been upgraded by the local technicians, the detailed information can found in the paper written by Bai et al.(2018).
Because of its filter system built on the primary focus, this can help to lessen optical loss and cut down the exposure time.
As a consequence, the integration time in $BVR_{c}I_{c}$ bands are 40s, 30s, 20s and 10s, respectively.
Another two stars near the target with similar brightness were chosen as the comparison star and the check star,
and the coordinates of them are listed in Table 1.

\begin{table}
\caption{Coordinates of NSVS 01286630,
the comparison star and the check star}
\centering
\begin{tabular}{|c|c|c|c|}
\hline
Stars             &  $\alpha_{j2000}$  &  $\delta_{j2000}$ & $V_{mag}$\\
\hline
NSVS01286630           & \emph{\emph{$18^{h}47^{m}08^{s}.59$}} &\emph{\emph{$+78^{\circ}42^{'}29^{''}.2$}}& 13.09\\
Comparison              & \emph{\emph{$18^{h}46^{m}05^{s}.37$}} &\emph{\emph{$+78^{\circ}38^{'}30^{''}.2$}}& 13.28\\
Check                   & \emph{\emph{$18^{h}47^{m}23^{s}.27$}} &\emph{\emph{$+78^{\circ}39^{'}46^{''}.3$}}& 12.76\\
\hline
\end{tabular}
\end{table}

All of observed CCD images were reduced with IRAF$^{1}$, including a flat-fielding and bias-fielding correction process.
\footnotetext{\small $1$ IRAF is distributed by the National Optical Astronomy Observatory, which is operated by the Association of the Universities for Research in Astronomy, inc. (AURA) under cooperative agreement with the National Science Foundation.}
By calculating the phase of the observations with a new equation, the observed LCs are plotted in Figure 1.
In order to avoid the phase shift, the phase of those observations displayed in Figure 1 were calculated with our new linear ephemeris:
\begin{equation}
  Min.(HJD)=2455504.0113(\pm0.00052)+0^{d}.38392787(\pm0.00000001)\times E.
\end{equation}
In addition, new CCD times of light minima for NSVS 01286630 were also observed and determined, which were listed in Table 2.
\begin{figure}
\begin{center}
\includegraphics[angle=0,scale=0.33 ]{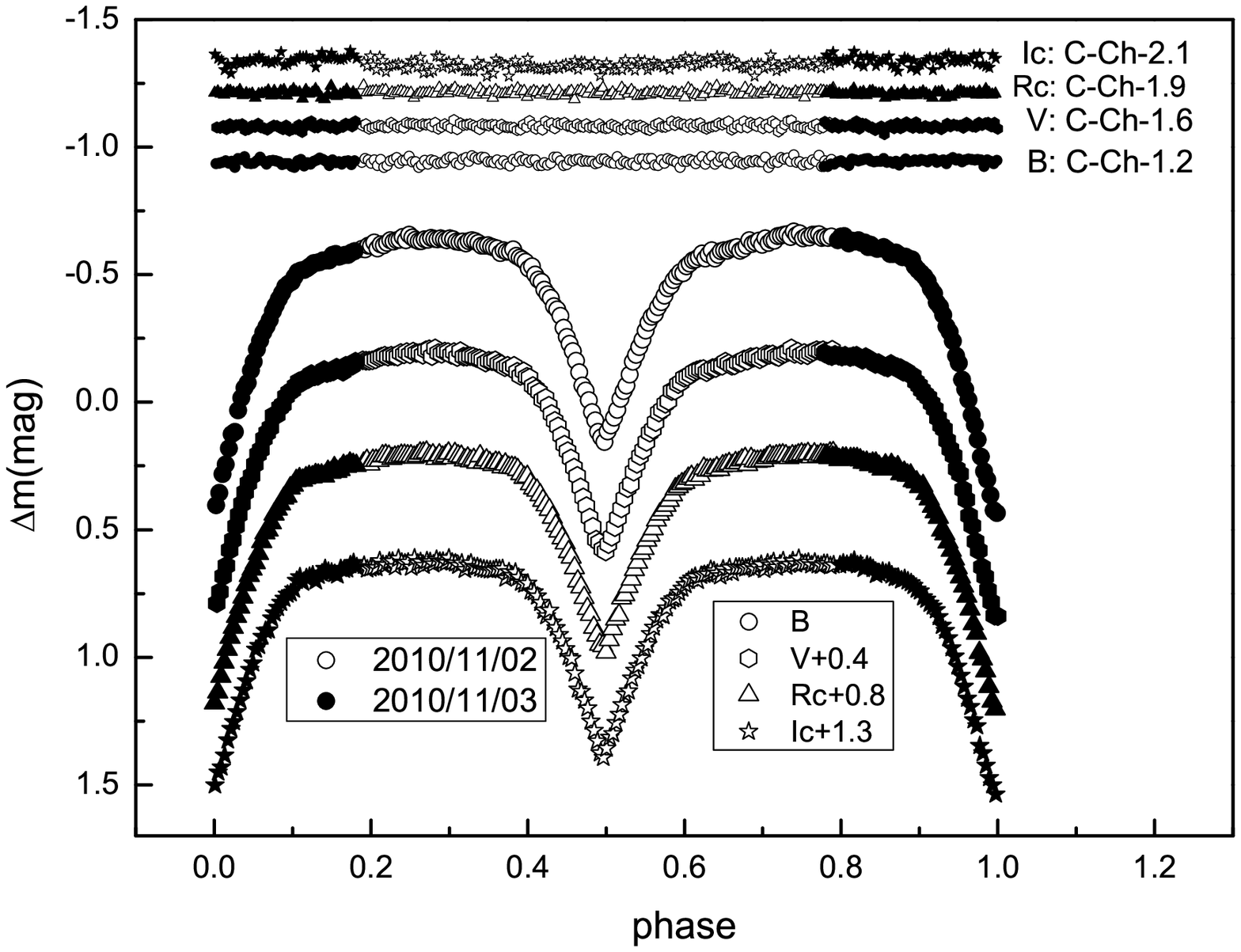}
\includegraphics[angle=0,scale=0.33 ]{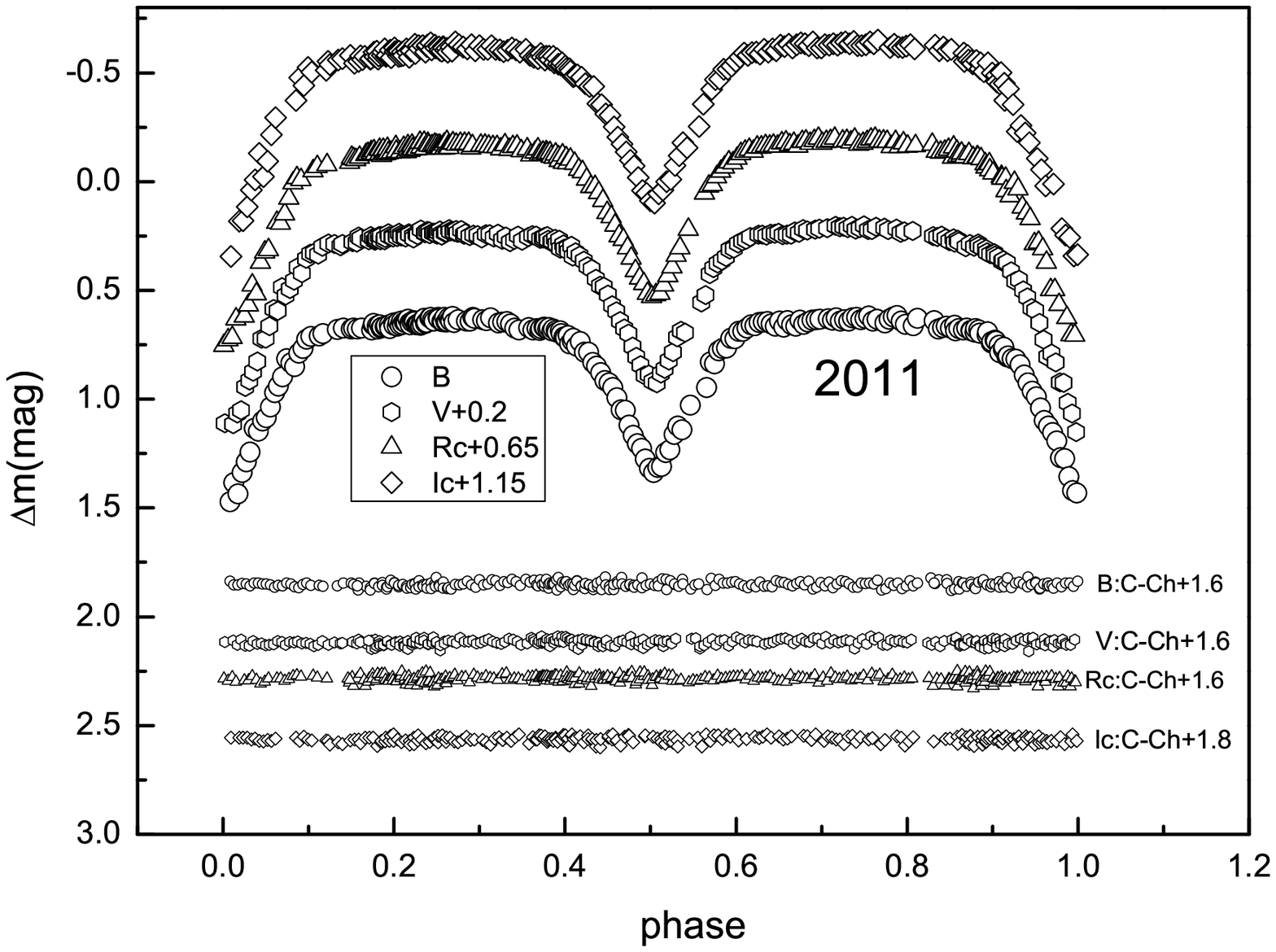}
\caption{ Light curves of NSVS 01286630 obtained in 2010 and 2011. Differential light curves of
the comparison star relative to the check star are plotted at the bottom marked $C-Ch$.}
\end{center}
\end{figure}

\begin{table}
\centering
\begin{minipage}{8cm}
\tiny
\caption{New CCD mean times of light minima for NSVS 01286630}
\begin{tabular}{rrll}\hline\hline
 JD(Hel.)    &  Error(d) &    Filter   & Telescope    \\
\hline
2455503.05012 & $\pm0.00040 $ &B      &85 cm\\
2455503.05015 & $\pm0.00034 $ &$I_{c}$&85 cm\\
2455503.05024 & $\pm0.00035 $ &$R_{c}$&85 cm\\
2455503.05010 & $\pm0.00040 $ &V      &85 cm\\
2455504.01103 & $\pm0.00039 $ &B      &85 cm\\
2455504.01142 & $\pm0.00061 $ &$I_{c}$&85 cm\\
2455504.01138 & $\pm0.00056 $ &$R_{c}$&85 cm\\
2455504.01137 & $\pm0.00050 $ &V      &85 cm\\
2455684.26348 & $\pm0.00043 $ &B      &85 cm\\
2455684.26361 & $\pm0.00049 $ &$I_{c}$&85 cm\\
2455684.26382 & $\pm0.00039 $ &$R_{c}$&85 cm\\
2455684.26245 & $\pm0.00041 $ &V      &85 cm\\
2455688.29501 & $\pm0.00018 $ &B      &85 cm\\
2455688.29517 & $\pm0.00024 $ &$I_{c}$&85 cm\\
2455688.29491 & $\pm0.00019 $ &$R_{c}$&85 cm\\
2455688.29509 & $\pm0.00025 $ &V      &85 cm\\
2455733.20970 & $\pm0.00063 $ &B      &85 cm\\
2455733.21128 & $\pm0.00045 $ &$I_{c}$&85 cm\\
2455733.21062 & $\pm0.00048 $ &$R_{c}$&85 cm\\
2455733.21239 & $\pm0.00057 $ &V      &85 cm\\
2455503.05040& $\pm0.00037$ &$BVR_{c}I_{c}$& 85 cm \\
2455504.01130& $\pm0.00052$ &$BVR_{c}I_{c}$& 85 cm \\
2455684.26334& $\pm0.00043$ &$BVR_{c}I_{c}$& 85 cm \\
2455688.29505& $\pm0.00022$ &$BVR_{c}I_{c}$& 85 cm \\
2455733.21100& $\pm0.00063$ &$BVR_{c}I_{c}$& 85 cm \\
\hline\hline
\end{tabular}
\end{minipage}
\end{table}

\section{Orbital Period Analysis of NSVS 01286630}
Long-time modern photometric observations for accumulating the minimum light times are necessary,
when using the $O-C$ method to analyze the period variation of EBs.
Generally, the short-period detached EBs with deep and symmetric eclipses,
which can help us to obtain individual minimum light times of mid-eclipses with high accuracy (Wolf et al. 2016).

The eclipsing binary NSVS 01286630 is one of new EBs with a precise orbital period of 0.38392787 days.
The eclipsing time changes of NSVS 01286630 were analyzed using all the minimum times published until now.
During our analysis, the following linear ephemeris were adopted
\begin{equation}
  Min.(HJD)=2454272.7532(\pm0.00008)+0^{d}.38392787(\pm0.00000001)\times E .
\end{equation}
There are altogether 91 CCD times of light minimum were used in our analysis, with 42 primary eclipses among them.
New CCD eclipse-times of NSVS 01286630 from our observations were determined
using the least-squares parabolic fitting method (Kwee \& van Woerden 1956; Li et al. 2017), whose uncertainties were determined by the error propagation formula.
The same minimum light times in different bands have been averaged, only the mean value are adopted.
Weights of 1/$\sigma^{2}$ were assigned to data, where $\sigma$ is error of the times of light minima.
We tried several different cases to fitting the $(O - C)$ diagram, including single sinusoidal function, parabolic add sinusoidal function.
To fit the $(O - C)$ diagram well, we considered a cyclic variation with an eccentricity of the orbital period.
By considering an eccentric orbit (Irwin. 1952), the fitting $(O-C)$ curve was described by the following equations:
\begin{eqnarray}
(O-C)&=&\Delta JD_{0} + \Delta P_{0}E\nonumber\\
 &&+A[\sqrt{1-e_{3}^{2}}\emph{\emph{sin}}E_{3}\emph{\emph{cos}}\omega_{3}+\emph{\emph{cos}}E_{3}\emph{\emph{sin}}\omega_{3}] ,\nonumber\\
\end{eqnarray}
\begin{eqnarray}
M_{3}&=&E_{3}-e_{3}\emph{\emph{sin}}E_{3}=\frac{2\pi}{P_{3}}(t-T_{3}) ,\nonumber\\
\end{eqnarray}
And in this case, the fitting residual is the smallest, most $(O - C)$ values can be well described using this new obtained equations,
hence, we taken it as the final solution.
More detailed explains and application about these equations used in present paper can found in the paper of Liao \& Qian (2010b).
Two possible mechanism can interpret this cyclical variation,
the light-travel time effect (LTTE) of the third body or a magnetic activity cycle for systems with an active late-type secondary star (Applegate 1992).
However, if the Applegate mechanism worked here, the mean value of period variation should be around 40-50 years for active EBs (Wolf et al. 2016).
Recently, the study for Kepler photometric database suggest that at least 8\% of close binaries have tertiary
companions with $P_{trip}$$\leq$ 7 years, which implies that multiplicity may be a common phenomenon among close EBs (Rappaport et al. 2013).
The apsidal motion effect in eccentric EBs can also result in a quasi-sinusoidal $(O-C)$ diagram, but, the formed $(O-C)$ curves
are anti-correlated between the primary minima and the secondary minima (Borkovits et al. 2016).
As discussed by Liao \& Qian (2010a), the presence of the tertiary component around the close binary systems should
mainly responsibility for that variations (Zhou et al. 2016). So, we think that the LTTE is the primary reason.

According to the $(O-C)$ fitting parameters, the orbital projected radius is calculated with the equation
\begin{equation}
  a_{12}\sin i_{3}=A_{3} \times c ,
\end{equation}
where $c$ is the speed of the light and $A_{3}$ is the amplitude of the $(O-C)$ oscillation,
i.e., $a_{12}$$\sin$$i_{3}$= 0.17($\pm$0.01) AU. The mass and the mass function of the tertiary
companion are computed with
\begin{equation}
  f(m)=\frac{4\pi^{2}}{GP_{3}^{2}}\times (a_{12}\sin i_{3})^{3}=\frac{(M_{3}\sin i_{3})^{3}}{(M_{1}+M_{2}+M_{3})^{2}} ,
\end{equation}
where $P_{3}$ is the period of the $(O-C)$ oscillation and $G$ is gravitational constant.
Finally, the best LTTE parameters and their errors are listed in Table 3.
From the Table 3, you can find that the mostly parameters we obtained are vary similar with the published results except orbital eccentricity. The main reason for that is the data adopted in the analysis, during our fitting, the data with errors more than 0.00005 days are not used, however,
about 112 reliable times of minimum light
are included in Wolfs' analysis, and some of them are not CCD measurements. 
It should be noted that the errors of these parameters are the unbiased standard errors, they
were calculated using the degree of freedom and the covariance matrix (Qian et al. 2015b).
The $(O-C)$ diagram and residuals for all data with respect to the above-mentioned linear ephemeris are shown in Figure 2.

\begin{figure}
\begin{center}
\includegraphics[angle=0.1,scale=0.33 ]{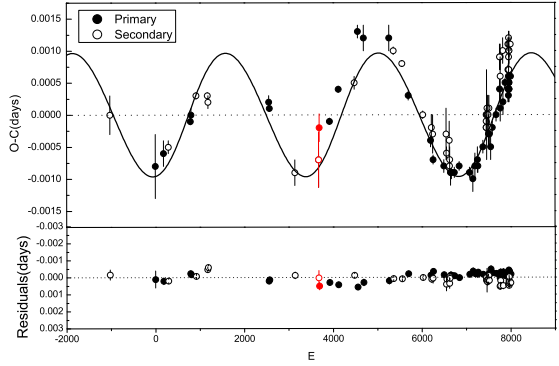}
\caption{ The $(O-C)$ diagram of NSVS 01286630 formed by all available
measurements with the linear ephemeris Equation (2). Open circles and dots refer to the
secondary and primary CCD  minimum
light times with error bars. The sine curve refers to the cyclic variation,
and the line refers to lineal variation. The corresponding
$(O-C)$ residuals are shown in the bottom panel.}
\end{center}
\end{figure}

\begin{table}
\tiny
\centering
\caption{The LTTE parameters for NSVS 01286630}
\begin{tabular}{lcll}\hline\hline
 Element &Unit     & Values (Present work)  & Values (Wolf et al. 2016) \\
\hline
Zero epoch, $T_{0}$                   & HJD & 2454272.75324($\pm$0.00005)   &2454272.75319($\pm$0.00008)   \\
Sidereal period, $P_{s}$              & days& 0.383927887($\pm$0.000000013) &0.383927874($\pm$0.000000012)  \\
Period of third body, $P_{3}$         & days& 1322($\pm$13)                 &1317($\pm$15)   \\
Eccentricity, $e_{3}$                 & -   & 0.08($\pm$0.10)               &0.17($\pm$0.05)   \\
$\omega_{3}$                          & deg & 342.23($\pm$1.3 )             &326.4($\pm$2.6 )   \\
The amplitude, A                      & days& 0.00097($\pm$0.00006)         &0.00105($\pm$0.00005)   \\
Periastron passage, $T_{3}$           & HJD & 2457144.34662($\pm$0.00010)   &2454450($\pm$15)  \\
$P_{s}^{2}/P_{3}$                     & days& 0.000115                      &0.000 11   \\
\hline
\end{tabular}
\end{table}
\section{Photometric Solutions with W-D Program}
The first photometric solutions of NSVS 01286630 in $V$, $R$ and $I$ bands were published in 2007.
In order to check these photometric elements and understand its evolution state,
we intend to analyse the present new LCs in the $BVR_{c}I_{c}$ bands with the 2013 version (W-D) code (Wilson \& Devinney 1971; Wilson 1979, 2012).
Before using the (W-D) code, we set the initial value for some fixed parameters.
The possible effective temperature for primary component (star 1), $T_{1}=4290 K$ was fixed according to average colour index.
The same value of gravity-darkening coefficients is taken, i.e., $g_{1}=g_{2}=0.32$ (Claret 2000), and 
the bolometric albedo for both components were sent as $A_{1}=A_{2}=0.5$ (Rucinski 1969).
The orbital eccentricity $e=0$, when the orbital period less than 1 days, the close binary orbits become circularized (Zahn 1989).
The adjustable parameters are: the mean effective temperature of the secondary component, $T_{2}$;
the mass ratio, $q$; the monochromatic light of star 1, $L_{1B}$, $L_{1V}$,
$L_{1R_{c}}$ and $L_{1I_{c}}$; the orbital inclination, $i$;
the dimensionless potentials of the two components $\Omega_{1}$ and $\Omega_{2}$.

As we all known, the reliable mass ratio of the EBs can be directly constraint by the high quality radial velocities (Becker et al. 2008).
The magnitude of the NSVS 01286630 in $V$-band is about 13$^{m}$.1 (Hoffman et al. 2008),
the fundamental parameters can be determined with high S/N ratio ($>$10) spectra, and this will require a telescope at least 4m class,
but, it is difficult for us now. Besides, because there is no spectroscopic observations for
NSVS 01286630 published up to now, we used a q-search method (fix $q$) to obtain initial input parameters.
During this process, we added a third light considering the fitting results of the $(O-C)$ diagram.
From Figure 1, we can see that the LCs observed in 2010 are of better quality, so we started the analysis with this set of LCs.
Using the grid computing method, a range from 0.5 to 3 (with a step of 0.1) was explored, however,
when the value of mass ratio less than 0.5 we can't find a convergent solution.
During the calculating, it is found that the solutions converged at mode 2 (for detached binaries) in the end.
The q-search result for NSVS 01286630 is shown in Figure 3, the optimal mass ratio we found to be $q=1.5$.
After that, we treated $q$ as a free parameter and set its initial value to be 1.5.
Then, one set of solution would derive when running the (W-D) code over and over again until all free parameters converged.
Although adopting the optimal mass ratio, the fitting results are still not good in its out-eclipses.
In view of this case, cool-star-spots model was adopted to fit the observed LCs (Zhang et al. 2018).
After many attempts, we found that one spot on the surface of the primary component
and the other one on the surface of the secondary component could fit the LCs pretty well, reducing the residual by 60\%.
It means that the two components of NSVS 01286630 are active stars covered with cool-star-spots in each other.
This case is similar with NSVS 11868841 whose study results are published several year ago (\c{C}akirli et al. 2010a).
At the same time, it should be noted that both of these cool-spots near the polar field.

According to the same process as stated above, we analyzed the LCs from 2011 using the (W-D) code again.
The similar spotted scenario are adopted in the finally solutions and the $\Sigma-q$ curves are generally same as the former.
However, the solutions are found to be have some different as the former, which can be caused by the shape changes of observed LCs.
Finally, after a detailed comparison, the best photometric solutions are listed in Table 4,
it should be noted that the listed errors calculated in (W-D) code are only probable errors (Liu et al. 2015).
These errors are obtained in the nonlinear situations, because of the complex correlations among the full parameters (Wilson \& Devinney. 1976),
the real parameter uncertainties maybe three to five times larger than the (W-D) code provided (Popper 1984; Abubekerov et al. 2009).
The theoretical LCs are plotted in Figure 4. The geometrical
structure of the system at different phases are displayed in Figure 5.
\begin{figure}
\begin{center}
\includegraphics[angle=0.1,scale=0.33 ]{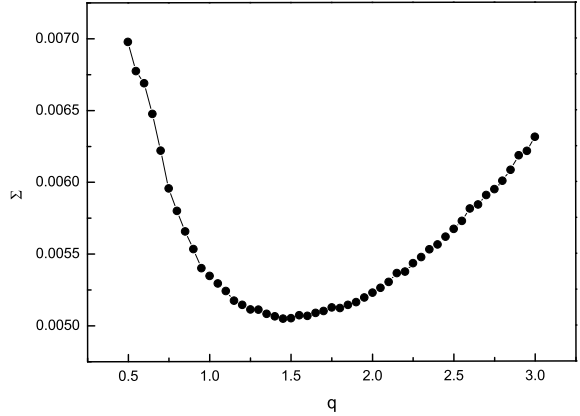}
\caption{The $\Sigma-q$ curves for NSVS 01286630 derived with $BVR_{c}I_{c}$-band light
curves in 2010. The optimal value of mass ratio is $q=1.5$.}
\end{center}
\end{figure}

\begin{figure}
\begin{center}
\includegraphics[angle=0,scale=0.33 ]{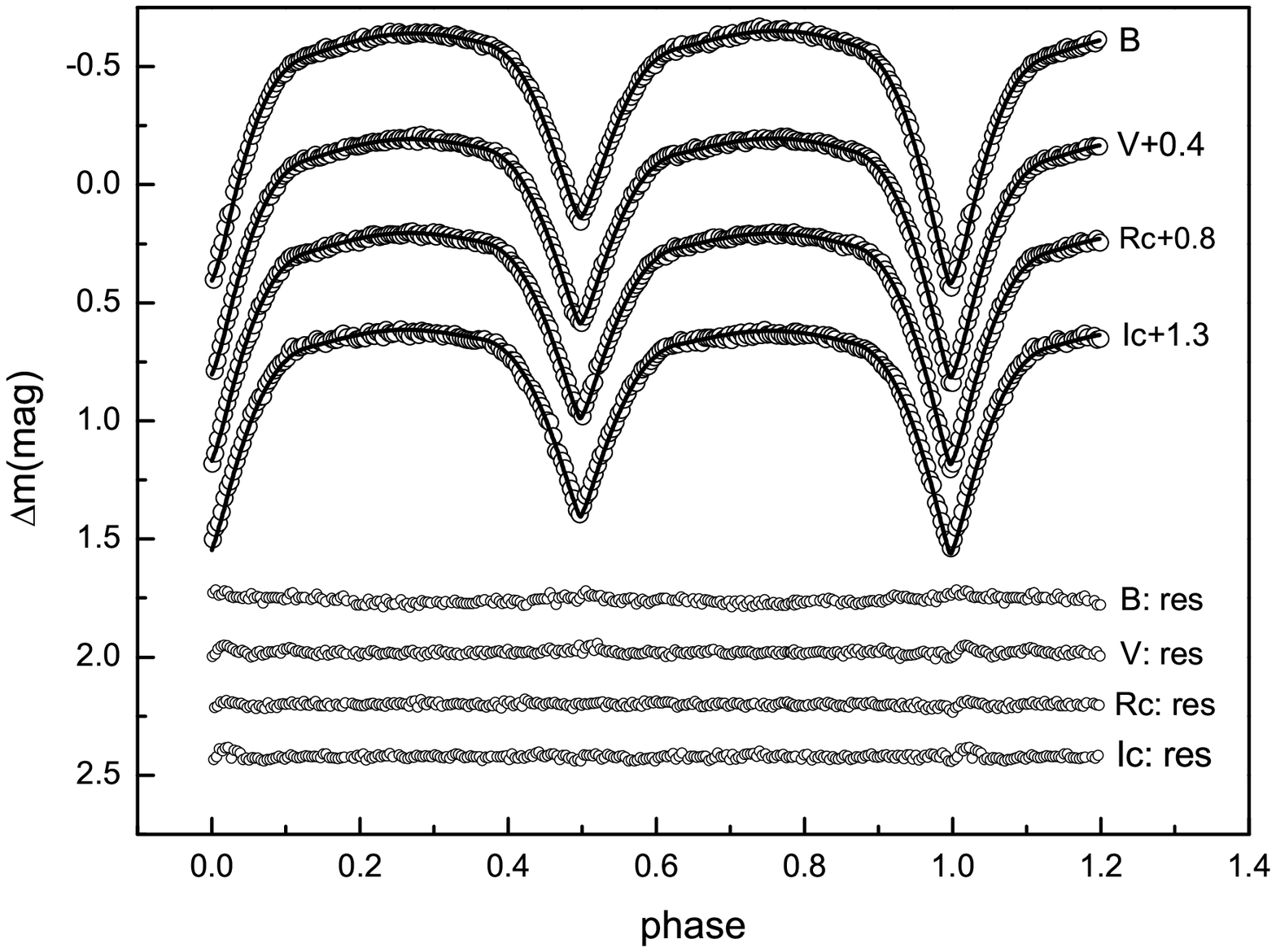}
\includegraphics[angle=0,scale=0.33 ]{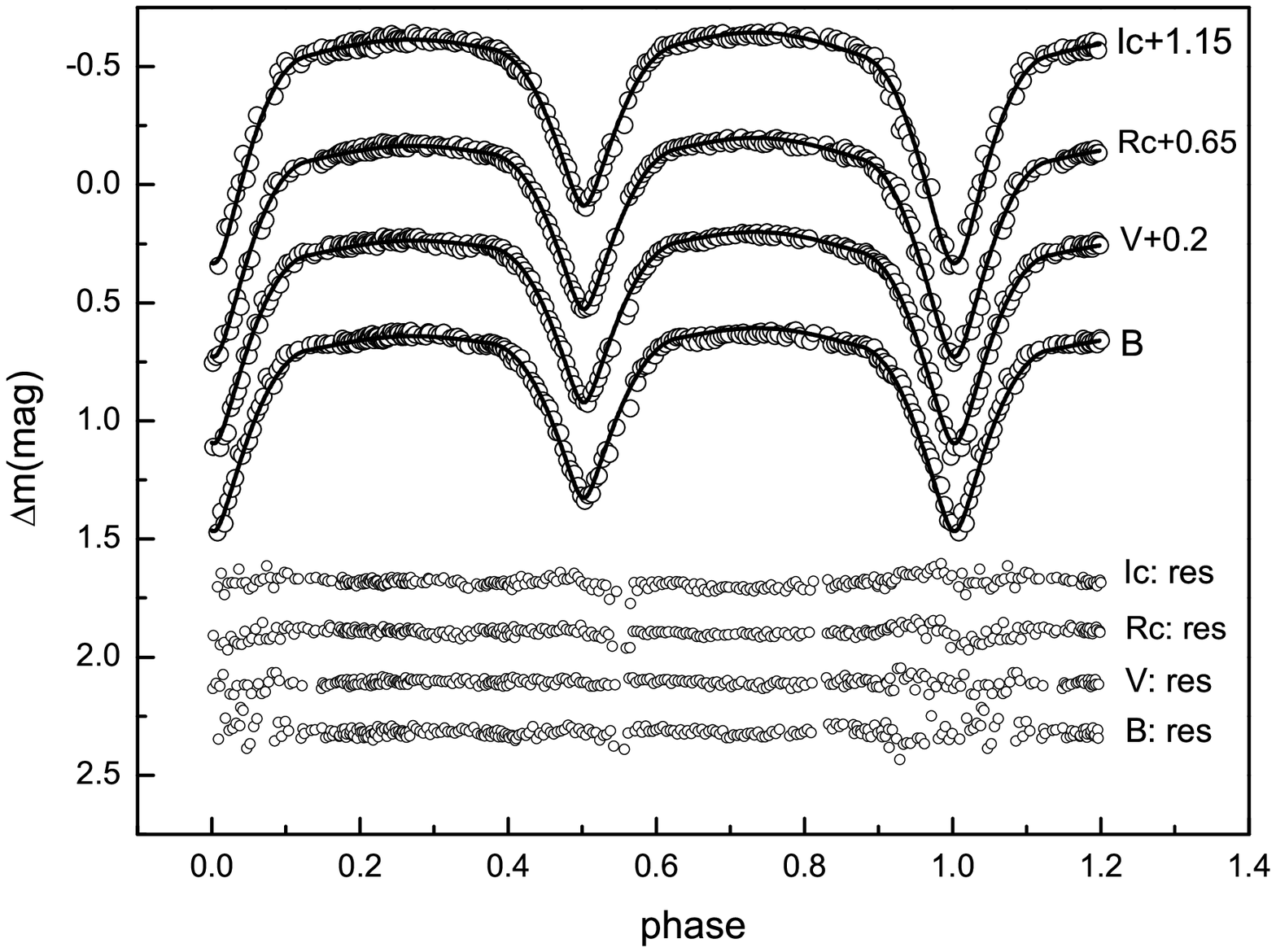}
\caption{The two sets of observational and theoretical LCs for NSVS 01286630 in $BVR_{c}I_{c}$ bands. The
different hollow symbols and black solid lines represent the observational and theoretical LCs, respectively.
Corresponding residuals between observed light curves and theoretical fits are plotted in the bottom.}
\end{center}
\end{figure}

\begin{figure}
\begin{center}
\includegraphics[angle=0.2,scale=0.33 ]{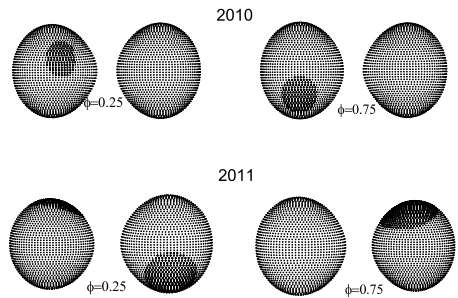}
\caption{Geometric configurations of NSVS 01286630 at different phases in 2010 and 2011.}
\end{center}
\end{figure}

\begin{table}
\tiny
\caption{Photometric solutions for NSVS 01286630}
\begin{tabular}{lll}
\hline
Parameters              &  This paper   &  This Paper \\
                        &   2010        &  2011  \\
\hline
$g_1=g_2$              &0.32                                         &0.32\\
$A_1=A_2$              &0.5                                          &0.5 \\
$T_1$(K)               & 4290                                        & 4290   \\
$q$                    & 1.12($\pm$0.10)                             & 1.15($\pm$0.15)    \\
$T_2/T_1$              & 0.9697($\pm$23)                             & 0.9655($\pm$58)    \\
$i$($^{\circ}$)        & 89.814($\pm$0.167)                          & 89.917($\pm$0.058) \\
$L_1/(L_1+L_2)(B)$     & 0.5678($\pm$0.0040)                         & 0.5509($\pm$0.0150)    \\
$L_1/(L_1+L_2)(V)$     & 0.5544($\pm$0.0035)                         & 0.5357($\pm$0.0121)           \\
$L_1/(L_1+L_2)(R_{c})$ & 0.5448($\pm$0.0038)                         & 0.5256($\pm$0.0122)           \\
$L_1/(L_1+L_2)(I_{c})$ & 0.5353($\pm$0.0042)                         & 0.5146($\pm$0.0126)          \\
$L_3/(L_all)(B)$       & 0.0048($\pm$0.0002)                         & 0.0044($\pm$0.0002)           \\
$L_3/(L_all)(V)$       & 0.0062($\pm$0.0002)                         & 0.0058($\pm$0.0002)           \\
$L_3/(L_all)(R_{c})$   & 0.0082($\pm$0.0003)                         & 0.0078($\pm$0.0003)           \\
$L_3/(L_all)(i_{c})$   & 0.0295($\pm$0.0004)                         & 0.0361($\pm$0.0006)          \\
$\Omega_1$             & 4.077($\pm$0.025)                           & 4.183($\pm$0.037)             \\
$\Omega_2$             & 4.204($\pm$0.030)                           & 4.201($\pm$0.028)            \\
$r_1(pole)$            & 0.3315($\pm$0.0010)                         & 0.3226($\pm$0.0038)             \\
$r_1(point)$           & 0.4026($\pm$0.0030)                         & 0.3826($\pm$0.0091)            \\
$r_1(side)$            & 0.3455($\pm$0.0012)                         & 0.3353($\pm$0.0044)            \\
$r_1(back)$            & 0.3700($\pm$0.0016)                         & 0.3574($\pm$0.0058)            \\
$r_2(pole)$            & 0.3375($\pm$0.0040)                         & 0.3455($\pm$0.0028)            \\
$r_2(point)$           & 0.3936($\pm$0.0052)                         & 0.4094($\pm$0.0068)             \\
$r_2(side)$            & 0.3509($\pm$0.0044)                         & 0.3602($\pm$0.0033)           \\
$r_2(back)$            & 0.3710($\pm$0.0060)                         & 0.3820($\pm$0.0042)            \\
$\theta_{s}$($^{\circ}$)   & P:292.574($\pm$2.349)                       &P:27.275($\pm$5.157)     \\
                           & S:129.90($\pm$2.865)                        &S:117.465($\pm$6.016)   \\
$\psi_{s}$($^{\circ}$)     & P:114.94($\pm$1.490)                        &P:71.338($\pm$4.240)      \\
                           & S:277.447($\pm$0.974)                       &S:94.946($\pm$1.489)   \\
$r_{s}$($^{\circ}$)        & 21.6($\pm$0.12), 24.87($\pm$0.17)           & 23.32($\pm$0.25), 24.92($\pm$0.28)    \\
$T_{s}/T_{\ast}$           & 0.86, 0.85                                  & 0.86, 0.85      \\
$\sum{(O-C)_i^2}$          & 0.00178                                     & 0.00410            \\
\hline
\end{tabular}
\begin{list}{}{  }
\item[Ref:]{ $L_{all}$ = $L_{1}$ + $L_{2}$ + $L_{3}$, $P$ refers to the primary component, $S$ means the secondary one. }
\end{list}
\end{table}

\section{Discussions and Conclusions}
First four-color light curve solutions for NSVS 01286630 are obtained using the 2013 version of the (W-D) code.
Due to the asymmetry of these LCs, a cool-star-spots model is used to analyze the LCs for better determination of basic orbital parameters.
The photometric solutions are relatively reliable considering an inclination nearly $90^{\circ}$.
The difference between two sets of photometric solutions may be caused by the evolution of the cool-star-spots.
The solutions from both sets of LCs suggest that the NSVS 01286630 is an active detached EB system with small temperature difference.
Besides, the photometric solutions show that the value of the mass ratio $q(\frac{M_{2}}{M_{1}})$ is more than 1, which is a very different
value compared with other late-type detached binaries (Torres \& Ribas 2002; Morales et al. 2009).
The main reason for this may be the presence of the big cool-star-spots on the secondary component of NSVS 01286630.
Our photometric solutions suggest that the cool-star-spots coverage on the surface of the secondary component is about 5\% (2010) and 11\% (2011), respectively.
Usually, the late-type stars with a deeper convective envelope and faster rotation, in turn, which will produce a strong magnetic field (Claret. 2000; Zhang et al. 2014).
Under the strong magnetic field and sufficient internal convection, the radius of these late-type low-mass stars
maybe inflation and the surface effective temperature going down at the same time (Chabrier \& Baraffe 2000; Morales et al. 2010).

The most striking feature of this binary system is its star-spots activity.
Therefore, the photometric solutions with cool-star-spots were adopted.
The surface cool-star-spots can make the observed LCs distorted, especially in short waves ($U-$, $B-$ and $V-$band), such as CU Cnc (Qian et al. 2012).
Generally, the cool-star-spots tend to occur around some longitude forming active longitude belts (Lanza et al. 2002; \c{C}akirli et al. 2003).
The main features of these longitudes are permanent active and continuously migrate (Berdyugina \& Tuominen 1998; Berdyugina 2005).
As previously emphasized that the cool-star-spots we obtained near polar regions, which are listed in the bottom of Table 5.
According to dynamo mechanism, late-type star's magnetic field is similar to magnetic dipole,
which owns a stronger magnetic field intensity near the polar regions than others considering its rotation (Durney \& Robinson  1982).
So, stars with fast rotators are more likely to exhibit cool-star-spots at high latitude and polar regions (Schuessler \& Solanki 1992).
More importantly, the tidal effects can result to the cool-stat-spots evolution along with the active longitudes and emerge on the surface (Zhang et al. 2014).
Tran et al. (2013) suggest that a cool-star-spot is continuously visible around the orbit and slowly changes its longitude on timescales of weeks to months.
It is the reason why the two-sets parameters of cool-star-spots we obtained are some different.
Similar late-type detached binaries include NSVS 02502726 (Lee et al. 2013); NSVS 10653195 (Zhang et al. 2015);
NSVS 6507557 (\c{C}akirli et al. 2010b) and NSVS 07453183 (Zhang et al. 2014).

The $(O-C)$ diagram of NSVS 01286630 shows a typical cyclical variation, which maybe caused by the presence of the third body or the magnetic activity from one or both components (Applegate 1992).
If the magnetic activity works here, the variations of the gravitational
quadrupole moment ($\Delta$Q) may result in the observed oscillations.
By using the formula of $\Delta$P/P = −9$\Delta$Q/Ma$^{2}$= 2$\pi$A/P$_{mod}$
(Lanza \& Rodon$\grave{o}$ 2002), where M is the mass of the active component, and
a is the separation between both components, respectively (Yang et al. 2012),
we can compute the values as follows: 
$\Delta$Q$_{1}$ = 9.64 $\times$ 10$^{48}$ g cm$^{2}$ 
and $\Delta$Q$_{2}$ = 1.14 $\times$ 10$^{49}$ g cm$^{2}$, which is evidently smaller than the
typical values of 10$^{51}$–10$^{52}$ g cm$^{2}$ for close binaries (Lanza \&
Rodon$\grave{o}$ 1999). Therefore, we can rule out Applegate’s mechanism
for interpreting the cyclic variations for NSVS 01286630 (Yang et al. 2014).
Hence, NSVS 01286630 maybe a triple system. Based on the fitting parameters from $(O - C)$ diagram, we can estimate the
 mass of the third body to be $M_{3}$sin($i_{3}$)=0.11 $M_{\odot}$, and the distance of the
binary system to the barycenter of triple system is calculated to be $a_{12}$sin($i_{3}$)= 0.17($\pm$0.01) AU.
In this case, the distance of third body to the barycenter of triple system can be calculated to be $a_{3}$sin($i_{3}$)=2.38 AU.
As yet, however, the data adopted in our analysis just spans 10 years,
so, further observations are required to ascertain the observed oscillation by us.

NSVS 01286630 is a detached close EB in which neither of components is filling the critical Roche lobe now.
It will evolve into a short-period W UMa-type binary until the components filled their critical Roche lobe.
So far, the formation and evolution of short-period W UMa-type binaries are still an unsolved problem in astrophysics,
NSVS 01286630 offers a good sample as DV Psc (Zhang \& Zhang 2007) to study it.
Some researchers (Liao \& Qian 2010a; Torres \& Ribas 2002) surveyed many EB systems with modern
photometric analysis and found that many of close binaries with additional companions. The spectral
type of these binaries ranged from $B$ to $M$, more than half of them are late-type EBs.
However, its origin is still questionable in several cases. One of theories suggest that
a intensive dynamic interactions process between binary system and these close-in stellar companions maybe change the formation and evolutional path of the stars.
For example, the third body might extract the angular momentum of the central binary system during the
early dynamical interaction or late evolution (Qian et al. 2014).
As a result, the third body may help to shorten the time of orbital evolution for these central binary systems (Fabrycky \& Tremaine 2007; Zhou et al. 2016).

\begin{acknowledgements}
This work is partly supported by Chinese Natural Science Foundation (No.11133007, 11573063 and 11503077 ),
the Science Foundation of Yunnan Province (grant No. 2012HC011).
We acknowledge the support of the staff of the Xinglong
85 cm telescope, and this work was partially supported by
the Open Project Program of the Key Laboratory of Optical Astronomy, National
Astronomical Observatories, Chinese Academy of Sciences.

\end{acknowledgements}

\label{lastpage}
\end{document}